\documentclass[seceq]{ptptex}
\usepackage{graphicx}
\usepackage{wrapft}

\title{
Hybridization effects and multipole orders in Pr skutterudites
}

\author{
Yoshio {\scshape Kuramoto}, 
Junya {\scshape Otsuki}, 
Annam\'{a}ria {\scshape Kiss}
and 
Hiroaki {\scshape Kusunose} 
}

\inst{
Department of Physics, Tohoku University, Sendai 980-8578, Japan
}

\abst{
Theoretical account is given of 4f-electron dynamics and multipole orders in Pr skutterudites with particular attention to 
(i) mechanism of the crystalline electric field (CEF) splitting leading to a pseudo-quartet ground state;
(ii) Kondo effect due to exchange interactions involving the pseudo-quartet;
(iii) multipole orders in the lattice of the pseudo-quartet in magnetic field.
Competition between the point-charge interaction and
hybridization between 4f and conduction electrons is identified as the key for controlling the CEF splitting.  
It is found that one of two pseudo-spins forming the pseudo-quartet has a ferromagnetic exchange, while the other has an antiferromagnetic exchange with conduction electrons. 
The Kondo effect is clearly seen in the resistivity calculated by the NCA, 
provided the low-lying triplet above the singlet is mainly composed of the $\Gamma_4$-type wave functions.
If the weight of the $\Gamma_5$-type is large in the triplet, the Kondo effect 
does not appear.
This difference caused by the nature of the triplet explains the presence of the Kondo effect in
PrFe$_4$P$_{12}$, and its absence in PrOs$_4$Sb$_{12}$.
By taking the minimal model 
with antiferro-quadrupole (AFQ) and ferro-type intersite interactions for dipoles and octupoles between nearest-neighbors,
the mean-field theory reproduces the overall feature of the multiple ordered phases in PrFe$_4$P$_{12}$. 
The AFQ order with the $\Gamma_3$-type symmetry is found to be stable only as a mixture of $O_2^0$ and $O_2^2$ components.
}

\begin{document}
\maketitle

\section{Introduction}
Novel many-body phenomena in solids often appear with the support of a specific structural feature of the system.  One of the typical examples is the high-temperature superconductivity in cuprates which is 
supported by the nearly two-dimensional square lattice. 
In this paper we discuss another interesting example of specific structural systems called filled skutterudites. 
A large number of compounds in this category 
are characterized by a cage-like structure of ligands in which rare-earth ions are loosely bound.
In filled Pr skutterudite compounds, 
many anomalous properties have recently been discovered 
such as the heavy-fermion superconductivity and high-field ordered phase in PrOs$_4$Sb$_{12}$,\cite{Bauer,Aoki} 
antiferro-quadrupole order in PrFe$_4$P$_{12}$,\cite{PrFeP} 
and metal-insulator and structural phase transitions in PrRu$_4$P$_{12}$. \cite{Sekine,Lee}
While neutron scattering has observed clear CEF transitions in PrOs$_4$Sb$_{12}$,\cite{Maple,Kohgi,kuwahara,gore} only broad quasi-elastic features are visible in PrFe$_4$P$_{12}$ above the temperature of quadrupole order \cite{Iwasa_Fe}.  
On the other hand, intriguing temperature dependence of CEF levels has been observed in PrRu$_4$P$_{12}$ below the metal-insulator transition \cite{Iwasa_Ru}.
Recently, a new phase 
has been found in PrFe$_4$P$_{12}$, \cite{tayama2} 
which appears only in high magnetic field along the (111) direction.
For proper understanding of these phenomena, 
both on-site and intersite interaction effects of 4$f$ electrons 
should be taken into account. 
The on-site hybridization with conduction electrons gives rise to CEF splittings in the second order and to the Kondo effect in higher orders.
The intersite interactions give rise to various electronic orders.

The purpose of the present paper is to give a comprehensive, though still crude, account of rich dynamical and ordering phenomena in Pr skutterudites.  
Particular attention is paid to the following aspects: 
(i) mechanism of crystalline electric field (CEF) splitting leading to a low-lying pseudo-quartet;
(ii) Kondo effect due to exchange interactions involving the pseudo-quartet;
(iii) multiple orders of the pseudo-quartet lattice in magnetic field.
Most of the theoretical results presented in this paper have been originally reported in refs.\citen{Otsuki1,Otsuki2,Kiss05}.   Here we organize these results toward a comprehensive picture.

\section{Mechanism of CEF splittings in Pr skutterudites}
\subsection{CEF parameters under the tetrahedral symmetry}
\begin{figure}
\centerline{
\includegraphics[width=0.4\textwidth]{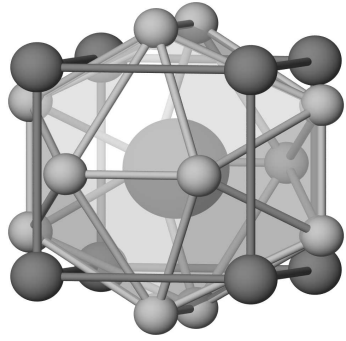}
}
\caption{
The cage structure forming a unit cell of filled skutterudites. .
}
\label{fig:structure}
\end{figure}

Filled skutterudites RT$_4$X$_{12}$ form a bcc structure with the space group $Im\bar{3}$.
In each unit cell, rare-earth ion R is surrounded by 8 transition metal ions T which form a cube, and 12 pnictogens X which form an icosahedron deformed slightly from the regular one.
Figure \ref{fig:structure} shows the cage structure with R in the center.
The R site has the local symmetry $T_h$ which has no four-fold rotation axis.\cite{THY}
In this symmetry, the $4f^2$ Hund's-rule ground states $^3H_4$ of Pr$^{3+}$ split into a singlet $\Gamma_1$, a 
non-magnetic doublet $\Gamma_{23}$, and two $\Gamma_4$ triplets.  
Of these, 
two $\Gamma_4$ triplets are written as $\Gamma_4^{(1)}$ and $\Gamma_4^{(2)}$, which are linear combinations of triplets $\Gamma_4$ and $\Gamma_5$ in $O_h$.
Under this crystal symmetry, the CEF potential is written as
\begin{align}
V_{\text{CEF}} 
&	 = A_4 [O_4^0+5O_4^4]+A_6^{\text{c}}[O_6^0-21O_6^4]+A_6^{\text{t}}[O_6^2-O_6^6]
\nonumber \\
& = W\left[ x \frac{O_4}{60} +(1-|x|)\frac{O_6^c}{1260}+y\frac{O_6^t}{30}
 \right],
\label{V_CEF}
\end{align}
in the standard notation\cite{THY}.
The term $yO_6^{t}$ mixes the $\Gamma_4$ 
and $\Gamma_5$  triplet states in the point group $O_h$.  

There are two main sources for the CEF splitting: 
the Coulomb potential from ligands, and hybridization between $4f$ electrons and ligands.
Let us begin with the Coulomb potential.
In the point charge model, CEF coefficients $A_4, A_6^{\rm c}$ and $A_6^{\rm t}$ in eq.(\ref{V_CEF}) are determined by coordination of the charge and the radial extension of the $4f$ wave function. 
By interpolating the data of ref.\citen{Freeman} for the radial extension,  we adopt the values 
$\langle r^4 \rangle =3.4a_{\rm B}^4$ and $\langle r^6 \rangle = 19 a_{\rm B}^6$ with $a_{\rm B}$ the Bohr radius.
Explicit results have been obtained in ref.\citen{Otsuki1} taking the effective charge $Z_{\rm t}$ of a transition ion as the parameter.
Here we require the charge neutrality
with trivalent Pr as
$$
3+4Z_{\rm t}+12Z_{\rm p}=0,
$$
where $Z_{\rm p}$ is the effective charge of a pnictogen ion.
The singlet $\Gamma_1$ is stabilized for $Z_{\rm t}>0$.
The stability of the singlet for positive $Z_{\rm t}$ implies that the maximum energy gain is provided by a more isotropic electron distribution associated with $\Gamma_1$.
The CEF potential becomes almost isotropic 
for $Z_{\rm t}\sim -0.25$.
The quasi-isotropy is due to high coordination numbers of pnictogens and transition ions, which tend to compensate the anisotropy.
For negative charge of transition ions, the electron distribution tends to be anisotropic to avoid the Coulomb repulsion.  In fact, 
the triplet $\Gamma_4^{(2)}$ is stabilized for $Z_{\rm t}<-0.25$.
To estimate the CEF potential, 
we use the lattice parameter in PrOs$_4$Sb$_{12}$\cite{Sugawara}, which gives
the Pr-T distance $d_{\rm t}=4.03\AA$ with T=Sb, the Pr-X distance $d_{\rm p}=3.48\AA$ with X=Os and the 
X-Pr-X vertical angle $2\theta_0=49.2 
^\circ$.
These data will be used in the next subsection to derive the results presented in Fig.\ref{fig:cef}.

\subsection{CEF splitting by p-f hybridization} 
Another important mechanism for CEF splittings is the covalency effect, or hybridization between localized and ligand orbitals.
According to band calculation,\cite{Harima} 
the conduction band striding  the Fermi level is formed mostly by the molecular orbital $a_u$ formed by 12 pnictogens around each Pr. 
The other relevant molecular orbital $t_u$ 
form two energy bands around a few eV above the Fermi level, and three a few eV below. 
We neglect contributions from $t_u$ bands to CEF splittings 
 because of the larger excitation energy. 
Therefore we concentrate on hybridization of the form
\begin{equation}
H_{\text{hyb}}
=V_{2u}\sum_{\sigma}f^{\dag}_{\sigma} c_{\sigma} + \text{h.c.},
\end{equation}
where $c_{\sigma}$ annihilates a conduction electron in the Wannier orbital with the $a_u$ 
symmetry at the origin,
and $f^{\dag}_{\sigma}$ creates an $4f$ electron with the same ($a_u$) orbital symmetry.
We take the hybridization parameter $V_{2u}$ real.
Hereafter we adopt the Mulliken notation such as $a_u$ for orbital symmetry, and the Bethe notation such as $\Gamma_1$ for a double-group representation with spin-orbit coupling.
In the second-order perturbation theory, the effective interaction is given by
\begin{equation}
	H_{\rm eff}= P H_{\text{hyb}} \frac{1}{E-H_0} Q H_{\text{hyb}} P,
\label{H_int}
\end{equation}
where $P$ is the projection operator onto 
$4f^2$ states, and $Q=1-P$.
We first deal with such part of the second-order hybridization that is diagonal with respect to the conduction states.
Diagonalization of $H_{\rm eff}$ with this constraint gives the 
CEF wave functions and their energies.  

We assume the following intermediate states:
(i) 4$f^{1}$ and an extra electron in vacant states, and
(ii) 4$f^{3}$ and extra hole in filled states.
For simplicity, we neglect the multiplet splittings in both kinds of intermediate states. 
Energy shifts in respective cases
are given by diagonalization of the following $(2J+1)$-dimensional matrices:
\begin{align}
\Delta E^{-}(M' ,M) &= -(1-n_{2u})
	 \frac{V_{2u}^2}
	 {\Delta_{1}+ \epsilon_{2u}} 
	 \sum_{\sigma}  
\langle M'| f^{\dag}_{ \sigma} 
	 f_{ \sigma} | M \rangle, \label{E-1} \\
\Delta E^{+}(M' ,M) &= -n_{2u}
	 \frac{V_{2u}^2 }
	 {\Delta_{3} + |\epsilon_{2u}|} 
	 \sum_{\sigma} 
\langle M'| f_{ \sigma} f_{ \sigma}^{\dag} | 
	 M \rangle, \label{E+1}
\end{align}
where 
$\Delta_{2\pm 1}\ (>0)$ are excitation energies to $4f^{2\pm 1}$, and 
$n_{2u}$ is filling of the pnictogen state $a_{2u}$ per spin. 
By adding contributions from $4f^{1}$ and $4f^{3}$ intermediate states, we obtain
\begin{align}
\Delta E(M' ,M) 
&= -\frac{V_{2u}^2}{\Delta_{3} + |\epsilon_{2u}|} 
2n_{2u} \ \delta_{M,M'} 
\nonumber \\ 
&+ V_{2u}^2 
\left[	 
\frac{ n_{2u}}{\Delta_{3} + |\epsilon_{2u}|} 
- 
\frac{1-n_{2u}}{\Delta_{1} + \epsilon_{2u}} 
\right] \sum_{\sigma} 
\langle M'| f^{\dag}_{ \sigma} f_{ \sigma} | M \rangle, 
\label{total}
\end{align}
where 
the first diagonal term does not contribute to the CEF splitting.
The matrix elements in eq.(\ref{total}) can be derived by explicitly using the $4f^1$ intermediate states as
\begin{align}
	\langle M'| f^{\dag}_{\sigma} | m_l m_{\rm s} \rangle
	= -\sqrt{2} \sum_{m_l' M_L M_S} \langle JM' | L M_L S M_S \rangle \nonumber \\
	 \times \langle L M_L | l m_l l m_l' \rangle
	 \langle S M_S | s m_{\rm s} s \sigma \rangle \langle l m_l' | l \Gamma_2 \rangle,
\label{matrix-el}
\end{align}
where $m,m_l$ denote azimuthal quantum numbers of $l=3$, and $m_{\rm s}$ denotes that of $s=1/2$.  The last factor in eq.(\ref{matrix-el}) describes the projection to the $a_{2u}$-type orbital of $4f$ electrons.  We obtain
$\langle l m_l | l \Gamma_2 \rangle = \pm 1/\sqrt{2}$ for $m_l=\pm2$,
and zero otherwise.

We note that  $4f^1$ and $4f^3$ intermediate states give opposite contributions to the level splitting.
Hence the sequence of CEF levels is determined by competition between both intermediate states\cite{Takahashi-Kasuya}.   
In eq.(\ref{total}), the competition appears in the factor with $n_{2u}$.
We have also done calculation where only Hund's-rule ground states in $4f^{2\pm 1}$ are considered as 
intermediate states.
The tendency of opposite contributions from $4f^1$ and $4f^3$ states persists, although the relative weight is not completely the same.  

The hybridization is parameterized by the Slater-Koster parameters $(pf\sigma)$ and $(pf\pi)$.\cite{Takegahara} 
It turns out hybridization with $a_u$ comes only from $(pf\pi)$,
which gives the interaction parameter   
\begin{equation}
V_{2u} = \frac 12\sqrt{30}(pf\pi) \sin2\theta_0 .
\end{equation}
With only the $a_u$ band taken into account,  eq.(\ref{total}) shows that the CEF level sequence is determined by two parameters; occupation of the $a_u$ band and $\Delta_3/\Delta_1$. 
To a good approximation we can assume the half-filled $a_u$ band, and
$\epsilon_{2u}=0$ for the band.
According to available information on PrP \cite{Takahashi-Kasuya},
we estimate the ratio roughly as $\Delta_3/\Delta_1=0.6$. 
Hence the level structure due to hybridization alone is more influenced by $4f^3$ intermediate states than by $4f^1$ ones.

\subsection{Combination of Coulomb interaction and hybridization}

\begin{figure}[htb]
\centerline{
\includegraphics[width= 0.7\textwidth]{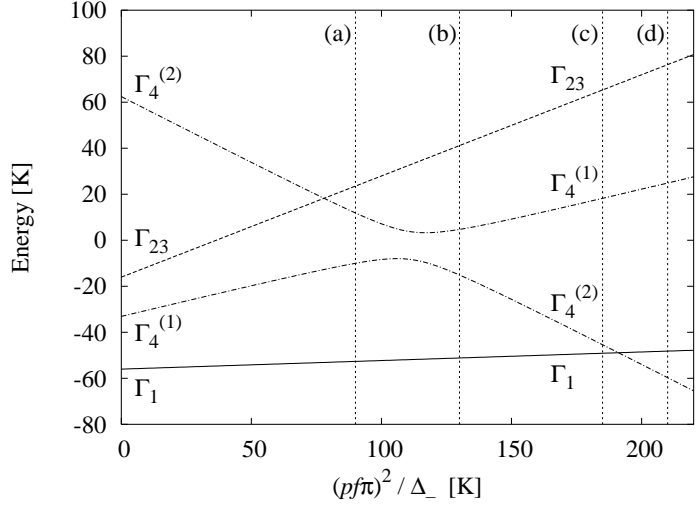}
}
\caption{
CEF level structures derived from hybridization and point charge potential as a function of 
$(pf\pi)^2/\Delta_-$. 
The level sequence qualitatively corresponds to: 
(a) PrRu$_4$P$_{12}$ in the high-temperature phase; 
(b) Pr1 site in PrRu$_4$P$_{12}$ in the low-temperature phase; 
(c) PrOs$_4$Sb$_{12}$;
(d) Pr2 site in PrRu$_4$P$_{12}$ in the low-temperature phase.  
See text for details.
}
\label{fig:cef}
\end{figure}

We  combine both contributions to CEF splittings by point-charge
interaction and by hybridization.  
Figure \ref{fig:cef} shows computed results 
as a function of strength of hybridization, $(pf\pi)^2/\Delta_-$ where we have introduced
$$
1/\Delta_-=1/\Delta_3 -1/\Delta_1.
$$
For the point charge parameters we tentatively take $Z_{\rm t}=2$ and $Z_{\rm p}=-11/12$.
We note that in M\"{o}ssbauer experiment on PrFe$_4$P$_{12}$, \cite{Mossbauer}
Fe is reported to be trivalent.  On the other hand, the energy-band calculation suggests a smaller valency for Fe\cite{Harima}.
The choice $Z_{\rm t}=2$ is a compromise between these conflicting results.
Overall CEF splittings become larger with larger $Z_{\rm t}>0$.
It should be emphasized that the the closely lying singlet and triplet are realized
only with simultaneous action of point-charge and hybridization interactions.

The lowest triplet $\Gamma_{\rm t}$, which is  either $\Gamma_4^{(1)}$ or $\Gamma_4^{(2)}$, can be written in terms of $O_h$ triplets $\Gamma_4$ and $\Gamma_5$ as
\begin{eqnarray}
|\Gamma_{\rm t},m\rangle
=\sqrt{w}|\Gamma_4,m\rangle+\sqrt{1-w}|\Gamma_5,m\rangle,
\end{eqnarray}
where $m=\pm, 0$ specifies a component, and $w$ with $0<w<1$ gives the weight of $\Gamma_4$ states.
We have chosen the phase of wave functions so that positive coefficients give the lower triplet.
As the parameter $(pf\pi)^2/\Delta_-$ increases, the $\Gamma_4^{(2)}$ state is stabilized and eventually crosses the $\Gamma_1$ level. 
After the crossing, the ground state acquires the local moment.
In the case of PrOs$_4$Sb$_{12}$,
the parameter $(pf\pi)^2/\Delta_- = 190$K is consistent with the combination of $\Delta_1 \simeq 5$eV, $\Delta_3 \simeq 3$eV and $(pf\pi) \simeq 0.35$eV.   
This value of $(pf\pi)$ is about 50\% 
larger than that estimated for PrP.\cite{Takahashi-Kasuya} 
Since the distance $d_{\rm p}\sim 3.5\AA$ in PrOs$_4$Sb$_{12}$
is smaller than the Pr-P distance 4.2$\AA$ in PrP, 
the larger magnitude is reasonable.
In the case of PrFe$_4$P$_{12}$, the lattice constant is 7.81$\AA$ as compared with 9.3$\AA$ of PrOs$_4$Sb$_{12}$.
Hence $(pf\pi)$ in PrFe$_4$P$_{12}$ should be even larger.

The level repulsion between $\Gamma_4^{(1)}$ and $\Gamma_4^{(2)}$ around $(pf\pi)^2/\Delta_-=110$K is due to mixing of wave functions, and is a characteristic feature in the point group $T_h$.   
In contrast,  a level crossing occurs in the $O_h$ group  
since there is no mixing between $\Gamma_4$ and $\Gamma_5$. 
Possible range of the effective hybridization in some representative Pr skutterudites is schematically shown by the lines (a)-(d) in Fig.\ref{fig:cef}.  In PrOs$_4$Sb$_{12}$, the triplet forming pseudo-quartet is almost of the $\Gamma_5$-type as shown by (c).  On the other hand, the CEF levels in the high-temperature phase of PrRu$_4$P$_{12}$ corresponds to (a), which bifurcates into (b) and (d) in the low-temperature phase.  Namely in one sublattice called Pr1, the triplet changes the character slightly, while in the other sublattice called Pr2, the triplet becomes the ground state.  Below the phase transition, the system behaves like an insulator.   It may be possible to describe the transition as the antiferro-hexadecapole order.  Explicit results of more detailed study will be reported elsewhere.
Finally, the case of PrFe$_4$P$_{12}$ should be mentioned.  As discussed later, the CEF levels seem to correspond to the range beyond the left end of Fig.\ref{fig:cef}.  
At first sight, the larger hybridization in PrFe$_4$P$_{12}$ seems incompatible with this leftward shift from PrOs$_4$Sb$_{12}$.
However, since the smaller lattice constant in PrFe$_4$P$_{12}$ favors the $4f^1$ excited states rather  than the $4f^3$ states,  the parameter $\Delta_1$ becomes smaller, leading to larger 
$\Delta_-$ and smaller $(pf\pi)^2/\Delta_-$.

\section{Exchange interaction in the pseudo-quartet}
\subsection{Second-order exchange interaction}

Now we deal with such components of $H_{\rm eff}$ given by eq.(\ref{H_int}), 
that give rise to exchange-type interaction between $4f$ and conduction electrons.
We restrict to the singlet and the lowest triplet for the CEF states for $4f$ states,
and neglect the multiplet splittings in the intermediate states.  
We obtain for $4f^1$ as intermediate states, 
\begin{align}
	H_{\rm eff} [4f^1]= \frac{V_{2u}^2}{\Delta_1} \sum_{MM'} \sum_{\sigma \sigma'}
	 A(MM' ; \sigma \sigma') \ |M \rangle  \langle M'|\ c^{\dag}_{\sigma} c_{\sigma'},
\label{exch4f1}
\end{align}
where $M$ denotes azimuthal quantum number of $J=4$, and
we have introduced the notation
\begin{equation}
	A(MM' ; \sigma \sigma') = 
\langle M| f^{\dag}_{\sigma'}
f_{\sigma}|M' \rangle,
\end{equation}
for the matrix element.
In the case of $4f^3$ intermediate states without multiplet splittings, 
the effective Hamiltonian is given by
\begin{align}
	H_{\rm eff} [4f^3] = -\frac{V_{2u}^2}{\Delta_3} \sum_{MM'} \sum_{\sigma \sigma'}
	 A'(MM' ; \sigma \sigma') \ |M \rangle  \langle M'|\ c^{\dag}_{\sigma} c_{\sigma'},
\end{align}
where 
$
	A'(MM' ; \sigma \sigma') = 
	\langle M| f_{\sigma}
f^{\dag}_{\sigma'} |M' \rangle. 
$
Because of completeness of the $4f^1$ and $4f^3$
intermediate states,
we obtain the relation
\begin{equation}
	A(MM' ; \sigma \sigma')=\delta_{MM'}\delta_{\sigma \sigma'}-A'(MM' ; \sigma \sigma'),
\label{f3tof1}	
\end{equation}
where the first diagonal term is irrelevant. 
Thus consideration of $4f^3$ intermediate states in addition to $4f^1$ 
is accomplished by the replacement
$$
V_{2u}^2/\Delta_1 \rightarrow V_{2u}^2(1/\Delta_1 + 1/\Delta_3)\equiv V_{2u}^2/\Delta,
$$
in eq.(\ref{exch4f1}).
Note that $4f^1$ and $4f^3$ states contribute additively in contrast with the case of CEF splittings. 
Although the bare diagonal terms are already included as CEF splittings, 
higher order potential scattering terms contribute to renormalization of  CEF levels as well as
damping of states.

\subsection{Symmetry analysis of the pseudo-quartet}

We are interested in the matrix element between the singlet and the triplet in the $T_h$ group.   As we have seen the triplet is a linear combination of $\Gamma_4$ and $\Gamma_5$ in the $O_h$ group.  
Hence it is convenient to analyze the symmetry properties of triplets in the cubic case.  
The direct products of pseudo-quartets can be decomposed as 
\begin{align}
	(\Gamma_1 \oplus \Gamma_4) \otimes (\Gamma_1 \oplus \Gamma_4)
	 &= \Gamma_1 + 2\Gamma_4 + (\Gamma_1 + \Gamma_3 + \Gamma_4 + \Gamma_5), \\
	(\Gamma_1 \oplus \Gamma_5) \otimes (\Gamma_1 \oplus \Gamma_5)
	 &= \Gamma_1 + 2\Gamma_5 + (\Gamma_1 + \Gamma_3 + \Gamma_4 + \Gamma_5),
\label{product}	 
\end{align}
where $\Gamma_4$ has the same symmetry as the magnetic moment. 
The product of 
$\Gamma_1\oplus\Gamma_4$ produces newly two $\Gamma_4$ representations, of which one is time-reversal odd and the other is even.
The odd representation represents the magnetic moment, while the even one the hexadecapole moment.
On the other hand, the pseudo-quartet $\Gamma_1 \oplus \Gamma_5$
 does not produce a new $\Gamma_4$ representation.
Physically, this means that $\Gamma_4$ as the first excited states gives rise to a van Vleck term in the magnetic susceptibility, while $\Gamma_5$ does not.

\begin{figure}[thb]
	\begin{center}
	\includegraphics[width=5cm]{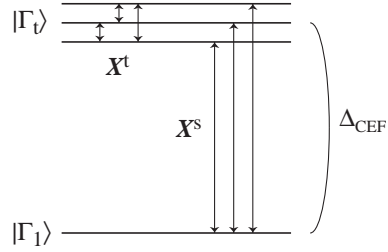}
	\end{center}
	\caption{Effective moment operators in the CEF singlet-triplet system.}
	\label{fig:x_operator}
\end{figure}

It is convenient to introduce the effective moment operators in the pseudo-quartet in $T_h$.  In the case of $w=0$ (pure $\Gamma_5$ triplet), 
the magnetic moment within $\Gamma_{\rm t}$ is the only relevant quantity.  The vector operator in this case is written as  $\mib{X}^{\rm t}$.
On the other hand, another vector operator 
$\mib{X}^{\rm s}$ is necessary to describe the magnetic moment of van Vleck type.  
The operator $\mib{X}^{\rm t}$ act on the triplet states, and $\mib{X}^{\rm s}$ connect the singlet and the triplet as shown in Fig. \ref{fig:x_operator}.

In terms of the four-dimensional basis set 
$|\Gamma_1 \rangle, |\Gamma_{\rm t} ,+ \rangle, |\Gamma_{\rm t} ,0 \rangle, |\Gamma_{\rm t} ,- \rangle$,
Each of the operators 
$X^{\rm t}_i , X^{\rm s}_j $ with $i,j =x,y,z$ can be represented by a $4\times 4$ matrix.  
Alternatively, one can use
the pseudo-spin representation as in ref.\citen{shiina-aoki}: 
$$
\mib{X}^{\rm t} = \mib{S}_1 + \mib{S}_2,\ 
\mib{X}^{\rm s} = \mib{S}_1 - \mib{S}_2,
$$
where $\mib{S}_1$ and $\mib{S}_2$ are spin 1/2 operators.
The pseudo-spin representation naturally leads to 
commutation rules among
$\mib{X}^{\rm t}$ and $\mib{X}^{\rm s}$. 

We  project the basis set $|M\rangle$ to the pseudo-quartet.
In terms of two vector operators 
$\mib{X}^{\rm t}$ and $\mib{X}^{\rm s}$, 
we then obtain the following form for the effective interaction within the pseudo-quartet:
\begin{align}
	H_{\text{exc}} &= 
	\left( 
	 I_{\rm t} \mib{X}^{\rm t} + I_{\rm s} \mib{X}^{\rm s}
	  \right)\cdot \mib{s}_c ,
\label{XtXs}	  
\end{align}
where 
$$
\mib{s}_c = \frac 12\sum_{\alpha\beta}
c^{\dag}_{\alpha} 
\mib{\sigma}_{\alpha \beta}c_{\beta} .
$$ 
The coupling constants are given by
$I_\alpha =V_{2u}^2 a_\alpha/\Delta$ for $\alpha=$ s,t
with
\begin{align}
	a_{\rm t}=-\frac{10+88w}{1155},\ a_{\rm s}=-\frac{4}{33} \sqrt{\frac{5w}{3}}.
\label{a_i}
\end{align}
In deriving eq.(\ref{a_i}), we have used eq.(\ref{matrix-el}).
Both $a_{\rm t}$ and $a_{\rm s}$ are negative, which means ferromagnetic exchange between conduction and $4f$ moments.  The origin of the ferromagnetic sign for $a_{\rm t}$ is traced to the Hund rule involving the spin-orbit interaction; the dominant orbital moment is pointing oppositely from the spin moment.  In other words, the spin exchange is antiferromagnetic as is usual for a hybridization induced exchange.  The sign of $a_{\rm s}$, on the other hand, is not physically meaningful.

\begin{figure}[tb]
\centerline{
\includegraphics[width=0.7\linewidth]{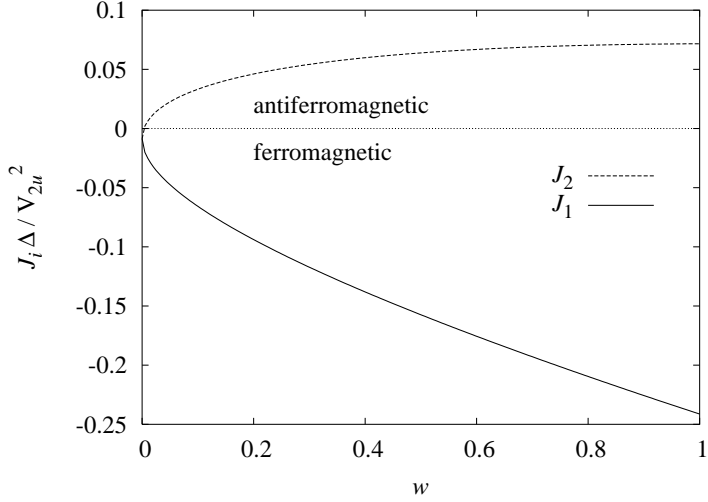}
}	
	\caption{Coefficients of pseudo-spins in the effective exchange.}
\label{c1c2}	
\end{figure}

The exchange interaction in terms of pseudo-spins is given by
\begin{equation}
	H_{\text{exc}} = 
	(J_1\mib{S}_1 + J_2\mib{S}_2)\cdot \mib{s}_c,
\label{pseudospin}	
\end{equation}
where 
$J_1= I_{\rm t}+ I_{\rm s}$ and 
$J_2= I_{\rm t}- I_{\rm s}.
$
Figure \ref{c1c2} shows the exchange interactions
$J_1$ and $J_2$ of pseudo-spins in units of $V_{2u}^2/\Delta$.
It should be noticed 
that $J_2$ becomes positive for $w > 0.00324$,  and represents the antiferromagnetic exchange.
The emergence of antiferromagnetic exchange is due to the particular CEF level structure in Pr skutterudites.  
The antiferromagnetic exchange 
is almost negligible in the pure $\Gamma_5$ case ($w=0$), and becomes an order of magnitude larger as $w$ increases toward unity, {\it i.e.}, toward pure $\Gamma_4$.  
It is likely that the Kondo-type behavior seen in PrFe$_4$P$_{12}$ originates from a large value of $w$ together with a small singlet-triplet splitting.
We shall next address to this problem.

\section{Kondo effect in the pseudo-quartet}
\subsection{Application of the NCA to $4f^2$ CEF states}
We proceed to derive dynamics of the singlet-triplet system by adopting the 
non-crossing approximation (NCA) \cite{nca1,kuramoto-kitaoka}.
In applying the NCA to systems with exchange interactions, we introduce a fictitious $4f^1$ intermediate state with negligible population\cite{bickers,maekawa}. 
We can alternatively introduce a fictitious $4f^3$ intermediate state instead of the $4f^1$ state with the same results. 
The NCA utilizes the resolvent $R_{\rm s}(z)$ for the singlet state and $R_{\rm t}(z)$ for the triplet.
The effects of the interactions are taken into account in terms of the self-energy $\Sigma_{\alpha}(z)$.
Each resolvent $R_{\alpha}(z)$ is given by
\begin{align}
	R_{\alpha}(z) = [z-\epsilon_{\alpha} - \Sigma_{\alpha}(z)]^{-1},
\end{align}
where $\epsilon_\alpha$ with $\alpha=$ s,t denotes CEF levels of $4f^2$ states, {\it i.e.},
$\Delta_{\rm CEF} = \epsilon_{\rm t} - \epsilon_{\rm s}$ without hybridization.
The NCA determines $\Sigma_{\alpha}(z)$ in a self-consistent fashion. 

There also appear renormalized exchanges $\tilde{I}_{\rm t}(z)$ and $\tilde{I}_{\rm s}(z)$, which are modified from the bare ones $I_{\rm t}$ and $I_{\rm s}$ in eq.(\ref{XtXs}). 
In addition, an effective potential $\tilde{K}_{\rm t}(z)$ for the triplet and $\tilde{K}_{\rm s}(z)$ for the singlet are generated by higher-order exchange scatterings. 
The self-energy is given in terms of the effective potential by the NCA integral equation:
\begin{align}
	\Sigma_{\alpha}(z) = -2 \int d\epsilon \rho_c(\epsilon) [1-f(\epsilon)]
	 \tilde{K}_{\alpha} (z-\epsilon),
\label{self}
\end{align}
where $\rho_c(\epsilon)$ is the density of states of the conduction band,
and $f(\epsilon)$ is the Fermi function with the chemical potential being 0.
In order to derive equations for the renormalized interactions $\tilde{I}_{\alpha}(z)$ and $\tilde{K}_{\alpha}(z)$, we divide the operator $\mib{X}^{\rm s}$ into two parts:
\begin{align}
	\mib{X}^{\rm s} = P_{\rm t}\mib{X}^{\rm s}P_{\rm s} + P_{\rm s}\mib{X}^{\rm s}P_{\rm t},
\end{align}
where $P_\alpha$ is the projection operator to the state $\alpha=$ s,t.
The first term operates on the singlet state, and second one to the triplet states.
Correspondingly, we define effective interactions $\tilde{I}_{\rm s}^{\rm (ts)}(z)$ and $\tilde{I}_{\rm s}^{\rm (st)}(z)$ for each part. 
The simultaneous equations for the effective interactions are given in the matrix form by
\begin{align}
	\left(
	\begin{array}{c}
		\tilde{I}_{\rm t} \\ \tilde{I}_{\rm s}^{\text{(ts)}} \\ \tilde{I}_{\rm s}^{\text{(st)}} \\
		\tilde{K}_{\text{t}} \\ \tilde{K}_{\text{s}}
	\end{array} \right) = \left(
	\begin{array}{c}
		I_{\rm t} \\ I_{\rm s} \\ I_{\rm s} \\ 0 \\ 0
	\end{array} \right) - \left(
\begin{array}{ccccc}
		I_{\rm t} \Pi_{\text{t}}/2 & 0 & I_{\rm s} \Pi_{\text{s}}/2 & I_{\rm t} \Pi_{\text{t}} & 0 \\
		0 & I_{\rm t} \Pi_{\text{t}} & 0 & 0 & I_{\rm s} \Pi_{\text{s}} \\
		I_{\rm s} \Pi_{\text{t}} & 0 & 0 & I_{\rm s} \Pi_{\text{t}} & 0 \\
		I_{\rm t} \Pi_{\text{t}}/2 & 0 & I_{\rm s} \Pi_{\text{s}}/4 & 0 & 0 \\
		0 & I_{\rm s} \Pi_{\text{t}} 3/4 & 0 & 0 & 0
	\end{array}
	\right) \left(
	\begin{array}{c}
		\tilde{I}_{\rm t} \\ \tilde{I}_{\rm s}^{\text{(ts)}} \\ \tilde{I}_{\rm s}^{\text{(st)}} \\
		\tilde{K}_{\text{t}} \\ \tilde{K}_{\text{s}}
	\end{array} \right),
\label{dyson_interaction}
\end{align}
where the auxiliary quantity $\Pi_{\alpha}(z)$ is  introduced by
\begin{align}
	\Pi_{\alpha}(z) = \int d \epsilon \rho_c (\epsilon) f(\epsilon) R_{\alpha}(z+\epsilon).
\label{pi}
\end{align}
Solving eq. (\ref{dyson_interaction}) for the renormalized interactions, we obtain
\begin{align}
	\tilde{I}_{\rm t} &= \frac{4(2I_{\rm t} - I_{\rm s}^2 \Pi_{\text{s}}) }
	 {(2-I_{\rm t} \Pi_{\text{t}})(4 + 4 I_{\rm t} \Pi_{\text{t}} - 3I_{\rm s}^2 \Pi_{\text{t}}\Pi_{\text{s}})},
\nonumber \\
	\tilde{I}_{\rm s} &\equiv \tilde{I}_{\rm s}^{\text{(ts)}} =  \tilde{I}_{\rm s}^{\text{(st)}}
	 = \frac{ 4I_{\rm s} }{(4 + 4 I_{\rm t} \Pi_{\text{t}} - 3I_{\rm s}^2 \Pi_{\text{t}}\Pi_{\text{s}})},
\nonumber \\
	\tilde{K}_{\text{t}} &= -\frac{4I_{\rm t}^2 \Pi_{\text{t}} + 2I_{\rm s}^2 \Pi_{\text{s}}
	 - 3 I_{\rm t} I_{\rm s}^2 \Pi_{\text{t}} \Pi_{\text{s}} }
	{(2-I_{\rm t} \Pi_{\text{t}})(4 + 4 I_{\rm t} \Pi_{\text{t}} - 3I_{\rm s}^2 \Pi_{\text{t}}\Pi_{\text{s}})},
\nonumber \\
	\tilde{K}_{\text{s}} &= -\frac{3 I_{\rm s}^2\Pi_{\text{t}}}
	{4 + 4 I_{\rm t} \Pi_{\text{t}} - 3I_{\rm s}^2 \Pi_{\text{t}}\Pi_{\text{s}}}.
\label{renorm_interaction}
\end{align}
The effective interactions $\tilde{I}_{\alpha}(z)$ are required 
for the dynamical magnetic susceptibility.  This quantity is computed explicitly in ref.\citen{Otsuki2} with due account of the so-called vertex correction \cite{nca4}.
In this paper, we only deal with the electrical resistivity for which the vertex correction is not necessary.

\subsection{Temperature dependence of resistivity}
We discuss how the CEF singlet together with the exchange interactions influences the resistivity.
First we note that the CEF splitting is renormalized significantly by fourth-order effects of hybridization, {\it i.e.}, the second order in the exchange interaction.  
By perturbation theory with respect to $I_{\rm s,t}$
we obtain the second-order shift
\begin{align}
	\Delta^{(2)}_{\rm CEF} = -D \rho_c^2 (I_{\rm s}^2 -I_{\rm t}^2)
	 = -D \rho_c^2 J_1 J_2,
	 \label{energy_shift}
\end{align}
which also appears as the difference 
$\Sigma_{\rm t}^{(2)}(z) -\Sigma_{\rm s}^{(2)}(z)$ of second-order self-energies.
The renormalization depends hardly on temperature and energy as long as $T, |z|\ll D$. 
Since we have $J_1J_2<0$, 
the second order exchange stabilizes the CEF singlet. 
On the other hand, higher order exchanges cause the Kondo effect.
The competition between the Kondo effect and the CEF effect depends on their characteristic energy scales. 
In computing dynamical quantities, we take $J_2=0$ to minimize renormalization of the CEF splitting. 
Even with $J_2=0$,  the pseudo-spin $\mib{S}_2$ interacts with conduction electrons indirectly through the other pseudo-spin $\mib{S}_1$. 
With the additional condition $\Delta_{\rm CEF}=0$, 
the indirect coupling disappears and 
the singlet-triplet Kondo model 
is reduced to the Kondo model with the pseudo-spin $\mib{S}_2$ being decoupled. 

In the NCA for the exchange interactions, the impurity $T$-matrix is given by
\begin{align}
	T(i \epsilon_n) = - \frac{1}{Z_f} \int_C \frac{dz}{2\pi i} e^{-\beta z}
	 \sum_{\alpha} \tilde{K}_{\alpha}(z) R_{\alpha}(z+i\epsilon_n),
	\label{t-matrix}
\end{align}
where $Z_f$ is the partition function of $4f$ electrons with account of interactions,
$\epsilon_n=(2n+1)\pi T$ is the Matsubara frequency of fermions, and the contour $C$ encircles all singularities of the integrand counter-clockwise. 
We utilize the two kinds of spectral intensities:\cite{nca3, kuramoto-kitaoka}
\begin{align}
	\eta_{\alpha}(\omega) &= - \frac{1}{\pi} \text{Im} R_{\alpha}(\omega + i\delta), \\
	\xi_{\alpha}(\omega) &= Z_f^{-1} e^{-\beta \omega} \eta_{\alpha}(\omega),
	\label{tilde_xi}
\end{align}
where $\delta$ is positive infinitesimal.
In a similar fashion,  we further define the spectral intensities $\eta_{\alpha}^{\rm (K)}(\omega)$ and $\xi_{\alpha}^{\rm (K)}(\omega)$ for the effective potential 
$\tilde{K}_{\alpha}(z)$, 
and $\eta_{\alpha}^{\rm (I)}(\omega)$, 
$\xi_{\alpha}^{\rm (I)}(\omega)$ for the renormalized exchange 
$\tilde{I}_{\alpha}(z)$. 
Actually the spectral functions $\xi_{\alpha}(\omega)$ are computed by another set of equations to avoid divergent Boltzmann factor at low temperature.\cite{nca3} 
Performing analytic continuation $i\epsilon_n \rightarrow \omega +i\delta$ to real frequencies in eq. (\ref{t-matrix}), we obtain
\begin{align}
	-\frac{1}{\pi} \text{Im}T(\omega + i \delta)
	= \sum_{\alpha} \int d\epsilon [\xi_{\alpha}^{(K)} (\epsilon) \eta_{\alpha}(\epsilon + \omega)
	 + \eta_{\alpha}^{(K)} (\epsilon) \xi_{\alpha}(\epsilon + \omega)].
\end{align}

\begin{figure}[thb]
	\begin{center}
	\includegraphics[width=0.7\textwidth]{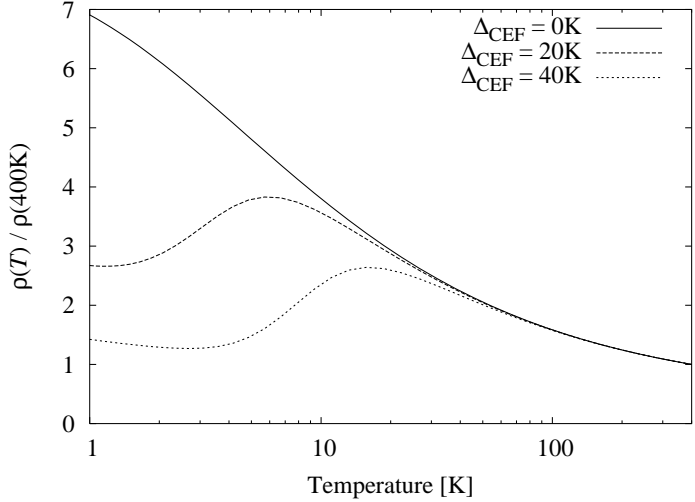}
	\end{center}
	\caption{The electrical resistivity as a function of temperature for selected values of $\Delta_{\rm CEF}$.}
	\label{fig:resistivity}
\end{figure}

The electrical conductivity $\sigma(T)$ 
is derived from $\text{Im}T(\omega)$ by 
\begin{align}
	\sigma(T)= A \int_{-\infty}^\infty d\epsilon \left( -\frac{\partial f(\epsilon)}{\partial \epsilon} \right)
	\frac{1}{|\text{Im}T(\epsilon)|},
\end{align}
where $A$ is a constant and $f(\epsilon)$ is the Fermi distribution function. 
We take a rectangular model with the band width $2D$ for conduction electrons
\begin{align}
	\rho_c(\epsilon) = \theta(D-|\epsilon|)/(2D),
\end{align}
where $\theta(\epsilon)$ is a step function. 
We take $D=10^4$K and use Kelvin as the unit of energy.
The cut-off energy $D$ is important in determining the Kondo energy scale, but otherwise does not enter the physics as long as it is much larger than other parameters.
Figure \ref{fig:resistivity} shows temperature dependence of the electrical resistivity $\rho(T)=1/\sigma(T)$ with $J_1\rho_c =0.2$ and $J_2=0$. 
In the quartet case $\Delta_{\rm CEF}=0$, $\rho (T)$ continues to increase as the temperature $T$ decreases.
On the other hand,  with the CEF singlet ground state
the enhancement of $\rho (T)$ 
is suppressed at temperatures below about the third of $\Delta_{\rm CEF}$. 
The suppression is due to a pseudo-gap of magnitude $\sim 2\Delta_{\rm CEF}$ in $\text{Im}T(\omega)$ around the Fermi level. 
Note that the temperature at the peak of $\rho(T)$ is substantially smaller than $\Delta_{\rm CEF}$. 
Namely the Kondo effect persists even though the CEF triplet is thermally depopulated.
This is understandable because the van Vleck-type polarization is responsible for the Kondo effect here.
Although the Kondo temperature is difficult to be defined unambiguously in this system, the magnetic relaxation rate at low $T$ is only a few K in the case of $\Delta_{\rm CEF}=0$.\cite{Otsuki2}

\section{Multiple orders in PrFe$_4$P$_{12}$}
\subsection{Splitting of the CEF triplet by 
magnetic fields}
We are now going to discuss the phase diagram in PrFe$_4$P$_{12}$ as a representative of the multiple ordered phases in magnetic field.
Let us compare the situation with PrOs$_4$Sb$_{12}$ where an AFQ order has been found only under magnetic field.
In the case of PrOs$_4$Sb$_{12}$, the CEF triplet consists mostly of $\Gamma_5$,  and the AFQ order also has the $\Gamma_5$-type symmetry.
In this case the splitting of the $\Gamma_5$ triplet under the molecular field is symmetric about the origin.  
Therefore, the quadrupole order parameters in A and B sublattices have the same magnitude.
In contrast, 
the splitting of the $\Gamma_4$ triplet under the molecular field is asymmetric about the origin in general.
This feature brings about interesting consequences in the AFQ order in PrFe$_4$P$_{12}$, which will be described below.

We begin with 
analysis of the CEF energy spectrum in magnetic fields 
taking the following single-site Hamiltonian:
\begin{equation}
H_{\rm ss}=V
_{\rm CEF}-g\mu_{\rm B} {\bf H}\cdot {\bf J} ,
\end{equation}
where the second term represents the Zeeman energy with the angular momentum {\bf J}, and 
the first term is the CEF potential given by eq.(\ref{V_CEF}).
The relevant parameter range for PrFe$_4$P$_{12}$
is $Wx<0$.\cite{Otsuki1}  
Then we obtain in the case of  $x>0$ and $W<0$,
\begin{eqnarray}
w=
\frac{1}{2}\left(1+\frac{3+2x}{\sqrt{(3+2x)^2+1008y^2}}\right), \label{weight}
\end{eqnarray}
and the CEF levels are given by 
\begin{eqnarray}
E(\Gamma_{\rm t})&=&
2W\left[x-4+2\sqrt{(3+2x)^2+1008y^2}\right]\,,\nonumber\\
E(\Gamma_1)&=& W\left(
108x-80
 \right)	\,.\label{g1g4}
\end{eqnarray}
The parameter $y$ 
can make the $\Gamma_1$ and $\Gamma_4^{(1)}$ states closely located to each other. 
For example, the choice $x=0.98$ and $y=-0.197$ gives the vanishing singlet-triplet splitting.
We recall that a constraint $|x|\leq 1$ is required in the parameterization of eq.(\ref{V_CEF}), and note that 
PrFe$_4$P$_{12}$ has the parameter is in the range $|x| \sim 1$.
Because of these reasons,  there appears apparent discontinuity in the CEF parameters as the CEF potential changes continuously.\cite{Otsuki1}  
In the case of $x <0$ and $W>0$, the weight $w$ and the CEF levels
can be obtained by the following replacement in eqs.(\ref{weight}) and (\ref{g1g4}):
\begin{equation}
x \rightarrow x/(2x+1), \ 
y \rightarrow y/(2x+1), \ 
W \rightarrow (2x+1)W.
\end{equation}
It is seen that $Wx$ and $Wy$ remains the same after the replacement.
In the special case of $x=-1$, the replacement above is just the reversal of signs for $x,y,W$.

\begin{figure}
\centering
\includegraphics[width=0.5\textwidth, angle=270]{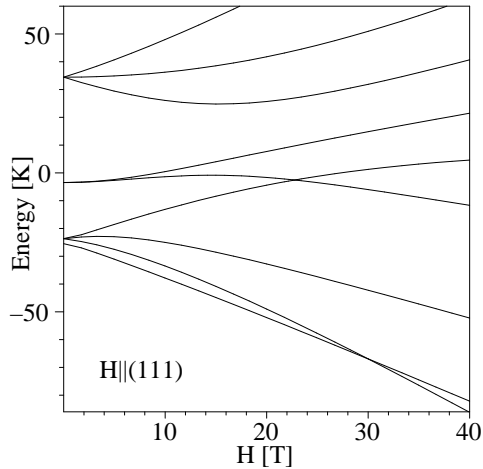}
\caption{Magnetic field dependence of the CEF states with $W=0.903$K, $x=-0.998$ and $y=0.198$ in the case of ${\bf H}\|(111)$.
The ground state at $H=0$ is the $\Gamma_1$ singlet with the $\Gamma_4^{(1)}$ triplet
located at 2K.} 
\label{fig:CEF111}
\end{figure}

\begin{figure}
\centering
\includegraphics[width=0.45\textwidth,angle=270]{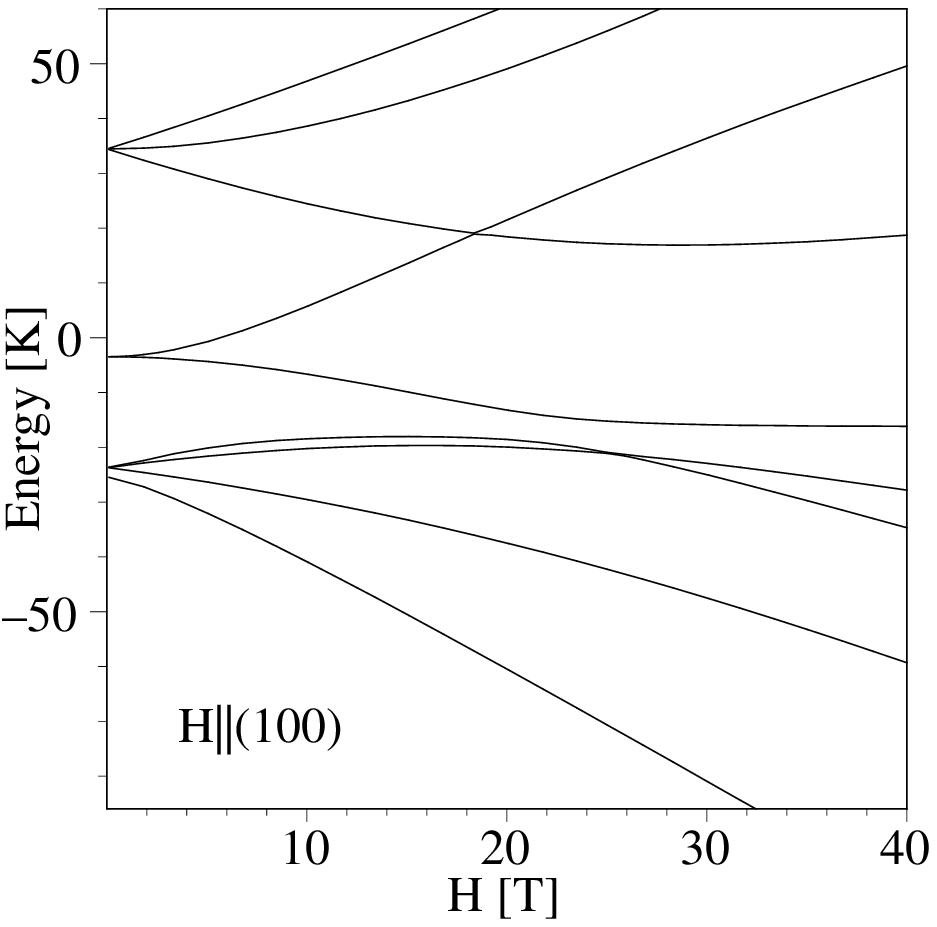}
\includegraphics[width=0.45\textwidth,angle=270]{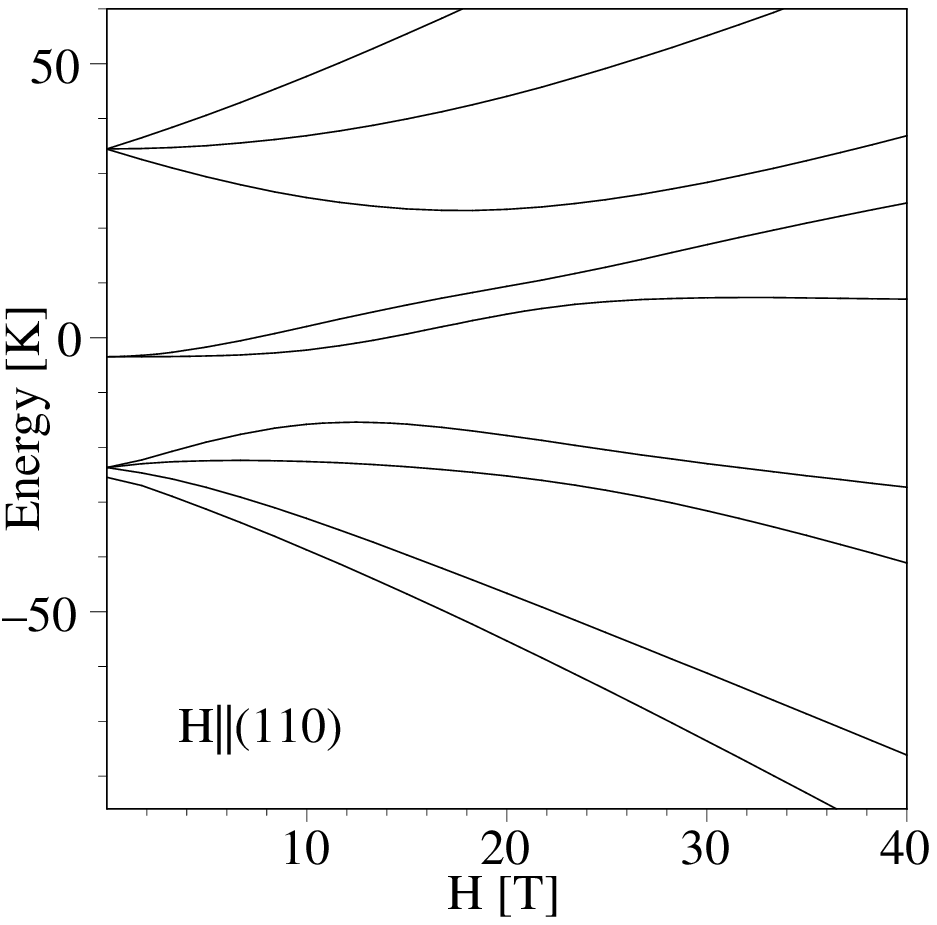}
\caption{Magnetic field dependence of the CEF states with the same parameters as in Fig.\ref{fig:CEF111}, but with ${\bf H}\|(100)$ (left) and ${\bf H}\|(110)$ (right).}
\label{fig:CEF2}
\end{figure}

Now we derive the eigenvalues of the $9\times 9$ matrix for $H_{\rm ss}$
as a function of magnetic field. 
Figure \ref{fig:CEF111} shows the result for magnetic field along (111) with the singlet-triplet splitting $\Delta_{\rm CEF}=2$K.
The most important feature is the level crossing near $H\sim 30$T, which was first noticed by Tayama et al.\cite{tayama2}
The level crossing occurs because of the level repulsion between the states originating from the doublet $\Gamma_{23}$ and the triplet $\Gamma_4^{(1)}$. 
Hence, if one neglects $\Gamma_{23}$ from the start, the level crossing does not occur.   The (111) direction of the magnetic field lowers the crystal symmetry to trigonal.   This symmetry is essential in bringing about the level crossing of the lowest two levels.  
For comparison, we have also derived the CEF spectrum for directions (100) and (110) of the magnetic field.  Figure \ref{fig:CEF2} shows the results.
It is evident that the lowest two levels never cross in these directions of the field.
We shall discuss later that the level crossing for (111)  is responsible for the ordered phase which appears only for this field direction, and for the anomalous angle dependence of the resistivity.\cite{Kiss05,kuramochi}

\subsection{Instability of the paramagnetic phase }
We now consider the simplest intersite interaction
that can reproduce the phase diagram of PrFe$_4$P$_{12}$
at least qualitatively.   
For this purpose we introduce notations for the quadrupole operators 
of the $\Gamma_3$-type, namely $\Gamma_{23}$ in the $T_h$ group, 
and such octupoles that reduce to $\Gamma_{5u}$ in the cubic symmetry. 
These are given by \cite{kubo-kuramoto04}
\begin{align}
{\bf  O_2} &=(O^0_2,  O^2_2) =\left( 
  \frac{1}{\sqrt{3}}(2J^2_z-J^2_x-J^2_y), \ 
  J^2_x-J^2_y \right) , \\ 
(T^{5u}_x,T^{5u}_y,T^{5u}_z) &= \frac{1}{2\sqrt{3}}
\left( 
\overline{J_x J^2_y}-\overline{J^2_z J_x},\ 
\overline{J_y J^2_z}-\overline{J^2_x J_y},\ 
\overline{J_z J^2_x}-\overline{J^2_y J_z}
 \right),
\label{T^5u}
\end{align}
where bars on the products represent normalized symmetrization,
\textit{e.g.}, 
$\overline{J_xJ^2_y} =(J_xJ^2_y+J_yJ_xJ_y+J^2_yJ_x)/3$.
We neglect the other quadrupole operators of the $\Gamma_5$-type in the intersite interaction.  This is because the induced antiferromagnetic moment in in PrFe$_4$P$_{12}$ is parallel to the magnetic field \cite{hao}, 
which suggests that the AFQ order is of $\Gamma_3$-type.
Then we take the following model for the intersite interaction:
\begin{align}
H_{\rm int} 
= 
\sum_{\langle ij\rangle}
\left( 
g_{\rm dip}{\bf J}_i\cdot {\bf J}_j  +
g_{\rm quad} {\bf O}_{2,i}\cdot {\bf O}_{2,j}+
g_{\rm oct}{\bf T}^{5u}_i \cdot {\bf T}^{5u}_j
\right),
\label{int3}
\end{align}
where $\langle ij\rangle$ means the nearest-neighbor pair,
and the coupling constants have signs $g_{\rm quad}>0, \ g_{\rm oct}<0$ and $g_{\rm dip}<0$. 
Namely we take antiferro-type quadrupolar and ferro-type dipolar and octupolar interactions. 
The presence of the ferromagnetic interaction is suggested by the temperature dependence of the magnetic susceptibility, and by the ferromagnetism occurring with La substitution \cite{aoki02}.
The dipole and $\Gamma_3$ quadrupolar interactions are consistent with the invariant form allowed by symmetry of the bcc lattice \cite{sakai}.
For simplicity we neglect the mixing term 
 ${\bf J}_i\cdot{\bf T}_j^{5u}$,
although the mixing is allowed by the $T_h$  symmetry.\cite{sakai}

We take the model given by $H_{\rm ss} +H_{\rm int}$,
and will obtain the phase boundary in the plane of magnetic field and temperature. 
The nine crystal field levels labeled by $| l\rangle$ 
are derived for different values of magnetic field by solving 
$H_{\rm ss} | l\rangle =E_{l}| l\rangle$. 
In deriving the phase boundary,  
we keep only the lowest six levels for simplicity of the numerical calculation. 
As seen from Fig.\ref{fig:CEF111},
the three higher levels neglected here repel with lower levels, and go up in energy for $H>20$T.
Thus their neglect will not influence the low-temperature behavior significantly. 

From the six levels labelled by $|l\rangle$ with $l=1,...,6$, 
we have $35\ (=6\times 6-1)$ independent pairs $k,l$ describing
multipole operators at each site. 
Among these, we have 15 symmetric combinations of the pairs, and
arrange them as increasing order of $10k+l$ with $k<l$.
Then we define $X^{\alpha}=| k \rangle \langle l |+| l \rangle \langle k |$ 
with $\alpha=1,...,15$ according to the ordering of the pair.
Similarly we define 15 antisymmetric combinations
$X^{\beta}=i\cdot(| k \rangle \langle l |-| l \rangle \langle k |)$ with $\beta =16,...,30$. 
The remaining five operators are diagonal in $k$ and $l$.
The multipolar interaction of eq.(\ref{int3}) is written as
\begin{eqnarray}
H_{\rm int}=-\sum_{\langle
ij\rangle}\sum_{\alpha,\beta}V^{\alpha\beta}_{ij}X^{\alpha}_{i}X^{\beta}_{j}
=-\sum_{\langle
ij\rangle} {\bf X}^{T}_{i}\cdot
{\hat{V}}_{ij}\cdot {\bf X}_{j}\,,
\end{eqnarray}
where ${\bf X}=[X^{1},X^{2},..,X^{35}]$, and $i,j$ are site indices. 
The susceptibility matrix in the momentum space
can be expressed as
\begin{eqnarray}
\hat{\chi}({\bf q})=\left(1-\hat{\chi}_0\cdot
\hat{V}({\bf q})\right)^{-1}\cdot
\hat{\chi}_0\,,\label{suscq2}
\end{eqnarray}
where 
$V^{\alpha\beta}({\bf q})$ 
is the Fourier transform of 
$V^{\alpha\beta}_{ij}$, 
and $\hat{\chi}_0$ is the single-site susceptibility matrix.   
The elements are given by 
\begin{eqnarray}
(\chi_0)^{\alpha\beta}=\sum_{k,l}\frac{\rho(E_{k})-\rho(E_{l})}
{E_{l}-E_{k}}\langle k |X^\alpha|l \rangle \langle l |X^\beta|k\rangle,
\label{nis}
\end{eqnarray}
where $\rho(E_{k})={\rm exp}(-\beta E_{k})/\sum_{k}{\rm exp}(-\beta E_{k})$.
In the right-hand side of eq.(\ref{nis}), we only need terms with $k\neq l$,
since the interaction Hamiltonian (\ref{int3}) does not have
the diagonal part of operators $X^{\alpha}$. 
The calculation of 
$(\chi_0)^{\alpha\beta}$ can easily be performed with use of the magnetic eigenstates, which are nondegenerate except at the level crossing point.

In the presence of ferromagnetic interaction, the effective field acting on CEF states can be stronger than the external magnetic field.
It is then important in deriving the phase diagram 
to include renormalization of the external magnetic field $h =g\mu_B H$.
We consider the effective magnetic field given by
$h_{\rm eff}=h-zg_{\rm dip}\langle J_{111}\rangle$
where the moment $\langle J_{111}\rangle$ along (111) 
depends on $h_{\rm eff}$ and temperature $T$.  
The effective field should be determined self-consistently.
Actually we first take $h_{\rm eff}$ as a given field and 
derive the matrix $\hat{\chi}_0$ under $h_{\rm eff}$.
To each value of $h_{\rm eff}$, we determine $h$ by the relation $h=h_{\rm eff}+zg_{\rm dip}\langle J_{111}\rangle$.

 susceptibility matrix. This is equivalent with the condition
\begin{eqnarray}
{\rm det}\left(1-\hat{\chi}_0\cdot
\hat{V}({\bf q})\right)=0.
\label{det}
\end{eqnarray}
The solution of eq.~(\ref{det}) gives the instability condition of the paramagnetic phase.  Note that the phase boundary may not coincide with the instability line if there is a first-order transition.
In this paper, we do not go into detailed inspection about the order of the transitions.

\subsection{Ordering vector and instability lines in magnetic field}

Under the reduced symmetry with finite magnetic field along the (111) direction,
the original $T_h$ system allows 
only two different symmetries for the 
order parameter:
the $C_3(\Gamma_1)$ singlet and the $C_3(\Gamma_3)$ doublet representations \cite{shiina-aoki}. 
Within the two-dimensional local Hilbert space, which is formed by the 
two states near the level crossing point at $H=H_{\rm cross}$, 
the order parameters with  $C_3(\Gamma_3)$  symmetry have nonzero inter-level matrix elements. 
Therefore, if the high-field phase has the same ordering vector as the low-field AFQ phase,  
both phases can be connected smoothly. 
If, on the other hand, the high-field phase has a different ordering vector ${\bf q}$,
the two phases cannot be connected smoothly.

\begin{figure}
\centering
\includegraphics[width=0.45\textwidth, angle=270]{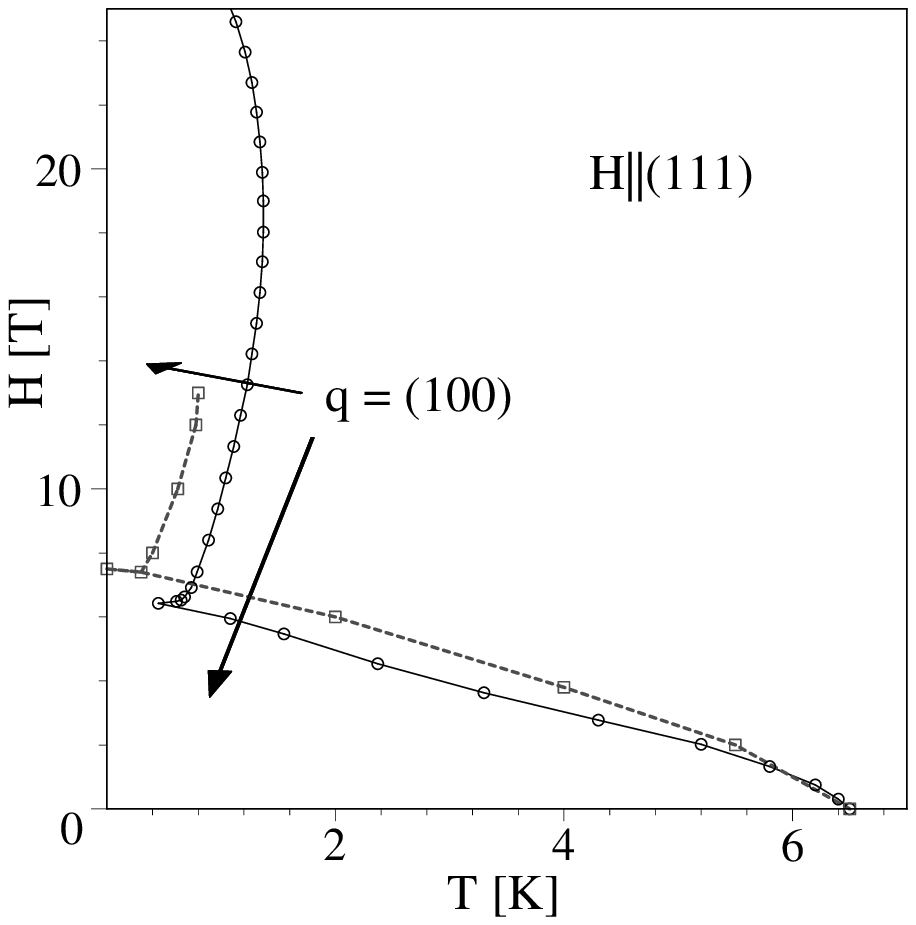}
\includegraphics[width=0.45\textwidth, angle=270]{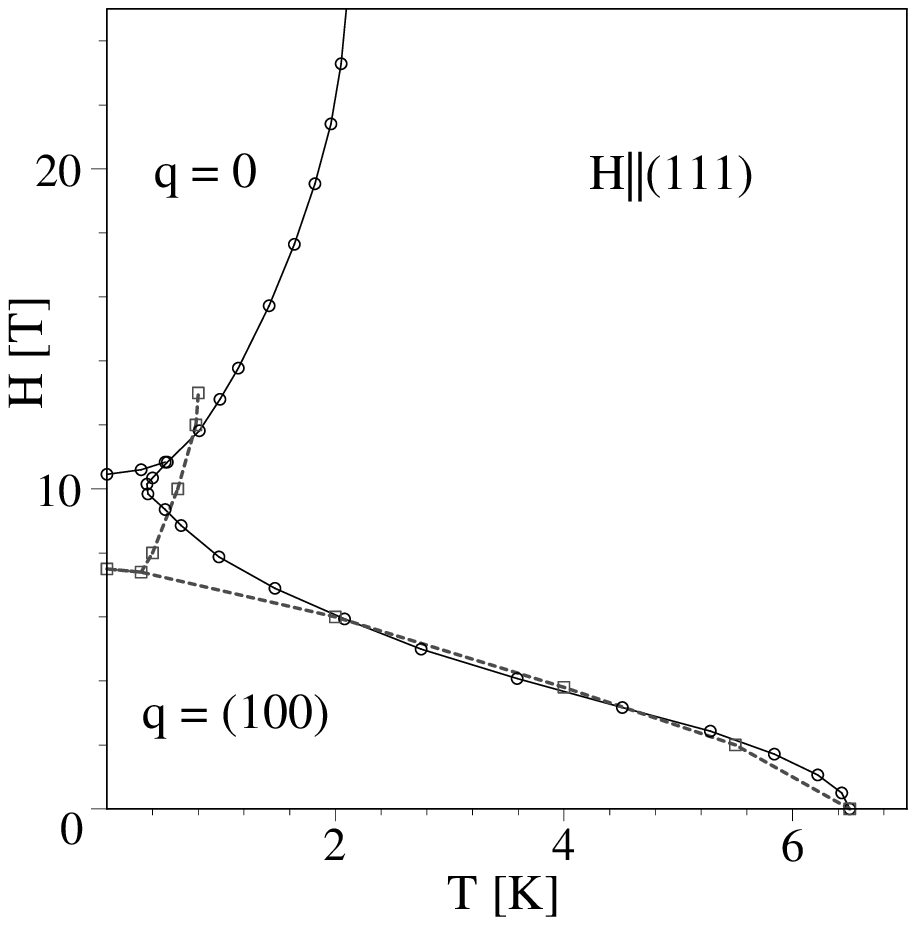}
\caption{Temperature--magnetic field phase diagram with two alternatives, 
${\bf q}=(111)$ or ${\bf q}=0$, 
 of the ordering vector in the high-field phase. Dashed line shows the observed phase boundary.  The parameters used are $g_{\rm quad}=8.93$mK for both left and right panels, and {\sl Left:}  
$g_{\rm dip}=-125$mK and $g_{\rm oct}=-25.7$mK; {\sl Right:} 
$g_{\rm dip}=-102.5$mK and $g_{\rm oct}=-34.2$mK.
}\label{fig:phases}
\end{figure}

We derive the instability 
from high temperatures toward the low-field and the high-field phases with several sets of interaction parameters.
We study both cases with the ordering vector ${\bf q}=(1,0,0)$, which leads to a two-sublattice order on the bcc lattice, and ${\bf q}=0$.
The quadrupolar coupling constant $g_{\rm quad}$ is determined to be 8.93 mK by the zero field transition temperature $T_Q=6.5$K. 
The small value of $g_{\rm quad}$ as compared with $T_Q$ is firstly due to the large numerical value of the quadrupole moment, and secondly due to the large value of $z=8$.
Figure \ref{fig:phases} shows examples of the calculated phase diagram
with the following parameters:
$W=-0.9$K, $x=0.98$ and $y=-0.197$, leading to $\Delta_{\rm CEF}=0$.
Two different types of solution can be obtained depending on the 
magnitude of ferro-type coupling constants: 
(a) both low- and high-field phases have the same 
ordering vector ${\bf q}=(1,0,0)$; 
(b) the high-field phase has the ordering vector ${\bf q}=0$. 
In the case (a) we obtain either continuous phase boundary or two separated phases.
For example, 
with $g_{\rm dip}=-125$mK and $g_{\rm oct}=-25.7$mK 
we obtain the continuous behavior as in the left panel in Fig.\ref{fig:phases},
while with the same $g_{\rm dip}$ but with $g_{\rm oct}=-25.9$mK, 
two phases are separated.
In the right panel, on the other hand, the high-field phase has the ordering vector ${\bf q}=0$.
In this case,  the the high- and low-field phases are separated by the first-order transition.
The effects of the 
ferro-type interactions on the phase boundary depend on the multipolar matrix elements and the 
mixing among multipoles due to the symmetry lowering as we increase the magnetic field. 
We found 
that the dipolar matrix elements are effective
in reduction of the transition temperature in the low-field regime ($H<H_{\rm cross}$), while
the $T^{5u}$ octupolar matrix elements have larger effects in the high-field regime ($H\sim H_{\rm cross}$). 
This latter behavior arises 
because the inter-level octupolar matrix elements between $\Gamma_4^{(1)}$--$\Gamma_{23}$ levels become significant 
as we increase the magnetic field. 
In the case of larger ferro-octupolar coupling, 
the high-field phase thus takes ${\bf q}=0$ as the ordering vector.

\subsection{Quadrupole patterns in the ordered phase}
In the presence of AFQ order, the relevant triplet states are split.
By application of magnetic field, staggered magnetic moments are induced in addition to the homogeneous moment.  
To analyze the splitting of the triplet,
we make use of  the $\Gamma_3$-type
quadrupolar moments 
$O_2^0$ and $O_2^2$ projected 
to the cubic $\Gamma_4$ or $\Gamma_5$ states.
In contrast with the $\Gamma_5$-type AFQ realized in PrOs$_4$Sb$_{12}$, the $\Gamma_3$-type AFQ does not involve the singlet $\Gamma_1$ as seen from eq.(\ref{product}).
Hence we concentrate here on the triplet only.
We write 
$|\Gamma_{\rm t}, m\rangle$  with $m=\pm,0$ simply as 
$|m\rangle$, and represent them as
$\{
|\alpha_1\rangle, |\alpha_2\rangle, |\alpha_3\rangle\}=\{
|0\rangle,|+\rangle,|-\rangle\}$.
Similarly the  cubic $\Gamma_4$ triplet states are represented by $|a_{i}\rangle$ ($i=1,2,3)$.
Then we define the $3\times 3$ matrices
${\cal O}^\mu_2$ and ${\cal C}^\mu_2$
with $\mu=0,2$ by 
\begin{eqnarray}
({\cal O}^\mu_2)_{ik} = \langle \alpha_{i}|O^\mu_2|\alpha_k \rangle, \ \ 
({\cal C}^\mu_2)_{ik} = \langle a_{i}|O^\mu_2|a_k \rangle. 
\end{eqnarray}
By comparing matrix elements in both basis sets, we obtain 
\begin{eqnarray}
{{\cal O}_2^0} &=& a_Q {\cal C}_{2}^0 + b_Q {\cal
C}_{2}^2\,,\nonumber\\
{{\cal O}_2^2} &=& a_Q {\cal C}_{2}^2 - b_Q {\cal
C}_{2}^0\,,
\label{g3k2}
\end{eqnarray}
where 
$a_Q=(9w-2)/7,\  b_Q=-\sqrt{3w(1-w)/7}$. 
The limits $w=1$ and 0 give the cubic
triplets $\Gamma_4$ and $\Gamma_5$, respectively.
Indeed for $w=1$, we recover 
${\cal O}_2^0={\cal C}_{2}^0$ and ${\cal O}_2^2={\cal C}_{2}^2$ 
from (\ref{g3k2})  because of $a_Q=1$ and
$b_Q=0$.  
The states $|0\rangle$, $(|+\rangle+|-\rangle)/\sqrt{2}$ and $(|+\rangle-|-\rangle)/\sqrt{2}$ are simultaneous eigenstates of
${\cal O}_2^2$ and ${\cal O}_2^0$.
The eigenvalues of ${\cal O}^\mu_2$ 
can be obtained from the eigenvalues 
$-14/\sqrt{3} (1,1,-2)$
of ${\cal C}_{2}^0$, 
together with eigenvalues $\pm 14$, $0$
of ${\cal C}_{2}^2$ by proper combination.
Generally, the eigenvalues of ${\cal O}^\mu_2$ consist of three singlets.
However, as shown below, 
the eigenvalues under the quadrupole molecular field consist of a doublet and a singlet.

Using eq. (\ref{g3k2}) we 
obtain 
\begin{eqnarray}
({\cal O}_{2}^0)^2+({\cal O}_{2}^2)^2 =
(a_Q^2+b_Q^2)[
({\cal C}_{2}^0)^2+({\cal C}_{2}^2)^2] = \frac{784}{3}(a_Q^2+b_Q^2),
\end{eqnarray}
which is a scalar matrix in consistency with the fact  that $(O_2^0)^2+(O_2^2)^2$
is a Casimir operator.
The factor $a_Q^2+b_Q^2 = (4-15d^2+60d^4)/49$ reduces to unity with $d=1$, and to 4/49 with $d=0$.  Namely the $\Gamma_5$ triplet has a smaller
quadrupole moment as compared with the $\Gamma_4$ triplet. 

In considering the AFQ
ordering pattern, it is convenient to 
begin with a two-site problem 
with sites $A$ and $B$ at zero temperature.
The interaction Hamiltonian is given within the $\Gamma_4^{(1)}$ subspace by
\begin{eqnarray}
{\cal H}(\Gamma_3)= \lambda_{\Gamma_3} ({\cal O}_{2,A}^{0}{\cal O}_{2,B}^{0}+{\cal O}_{2,A}^{2}{\cal O}_{2,B}^{2}). \label{tsham1}
\end{eqnarray}
We take the nine-fold basis $| k\rangle_A | l\rangle_B$ for the two sites, where $k,l$ can be $0$, $+$ or $-$.
By diagonalization of ${\cal H}(\Gamma_3)$
we find
that the nine-fold degeneracy splits into a six- and a three-fold multiplets. With antiferro-type interaction ($\lambda_{\Gamma_3}>0$), the ground state is the six-fold multiplet with states
\begin{eqnarray}
&&|0\rangle_A |+\rangle_B, \ \ 
|0\rangle_A |-\rangle_B, \ \ 
|+\rangle_A |0\rangle_B, \ \ 
|-\rangle_A |0\rangle_B, \nonumber\\
\frac{1}{\sqrt{2}} ( |+\rangle_A |+\rangle_B-|-\rangle_A |-\rangle_B)\label{tssstates}.
\end{eqnarray}
To obtain insight into the nature of degenerate wave functions,  we associate the triplet states with $l=2$ spherical harmonics as
\begin{eqnarray}
|0\rangle \sim  Y_2^0 \sim 3z^2-r^2, \ \
 |+\rangle \sim  Y_2^{2} \sim (x+iy)^2, \ \
 |-\rangle \sim  Y_2^{-2} \sim (x-iy)^2.\label{sphh}
\end{eqnarray} 
If we choose the state $|0\rangle_A$, namely $(3z^2-r^2)_A$, 
the six-fold degeneracy allows combinations 
$|+\rangle_B \pm |-\rangle_B$ to make 
$(3z^2-r^2)_A(x^2-y^2)_B$ and $(3z^2-r^2)_A(xy)_B$ as the degenerate ground states.
Similarly, combinations of states in the second line of eq.(\ref{tssstates})
give other degenerate ground states $(xy)_A(x^2-y^2)_B$ and  $(x^2-y^2)_A(xy)_B$. 
We remark that the six-fold degeneracy comes from the number of ways to choose different orbitals at two sites.  

With 
$|0\rangle$, $(|+\rangle+|-\rangle)/\sqrt{2}$ and $(|+\rangle-|-\rangle)/\sqrt{2}$ 
chosen as the basis at each site, 
the mean-field theory at zero temperature becomes exact for the two-site system.
The absence of off-diagonal elements in the two-site Hamiltonian 
persists in the bcc lattice, which can be separated into $A$ and $B$ sublattices.
Hence with only the $\Gamma_3$ intersite interaction, the degenerate ground state of the bcc lattice should be given exactly by the mean-field theory.
Recognizing this situation, we
consider the mean field of ${\cal H}(\Gamma_3)$ at site $B$:
\begin{eqnarray}
{\cal H}_{B}=\lambda_{\Gamma_3} \left ( \langle {\cal O}_{2}^{0}\rangle_{A} {\cal O}_{2}^{0}+\langle {\cal O}_{2}^{2}\rangle_{A} {\cal O}_{2}^{2}\right ) \equiv 
\lambda_{\Gamma_3} \left ( Q_{A} {\cal O}_{2}^{0}+q_{A} {\cal O}_{2}^{2}\right ) ,
\label{tssham2}
\end{eqnarray}
where we have used the notations $\langle {\cal O}_{2}^{0}\rangle_A \equiv Q_A$ and $\langle {\cal O}_{2}^{2}\rangle_A \equiv q_A$.
If we choose the state $|0\rangle_A$, 
${\cal H}_{B}$ is diagonal 
and has the following structure:
\begin{eqnarray}
h_{B0}\equiv \frac{1}{\lambda_{\Gamma_3}}\{
\langle \alpha_k|{\cal H}_B| \alpha_l\rangle
\}
&=& 
2 c   \left( \begin{array}{ccc}
                2  & 0 & 0\\
          0  & -1 & 0\\
          0  & 0 &  -1
     \end{array}
       \right), \label{matr1}
\end{eqnarray}
where $c= \left(14/\sqrt{3}\right)^2  (a_Q^2+b_Q^2)$ and we have used   $Q_A=28/\sqrt{3}a_Q$ and $q_A=-28/\sqrt{3}b_Q$. 
We note in eq.(\ref{matr1}) that 
the splitting into a singlet and a doublet is due to 
the mixing of moments ${\cal O}_2^2$ and ${\cal O}_2^0$ in the quadrupolar mean field. 
This quadrupolar field can be expressed as
\begin{eqnarray}
Q_{A} {\cal O}_{2}^{0}+q_{A} {\cal O}_{2}^{2}=28/\sqrt{3}(a_Q^2+b_Q^2){\cal C}_{2}^0\sim 3z^2-r^2.
\end{eqnarray}
The states $|\pm\rangle_B$ constitute the degenerate ground state.
If we choose
the states 
$(|+\rangle\pm |-\rangle)_A$, on the other hand,
the resultant quadrupolar fields are now given by
\begin{eqnarray}
&& Q_{A} {\cal O}_{2}^{0}+q_{A} {\cal O}_{2}^{2} = -14/\sqrt{3}(a_Q^2+b_Q^2)({\cal C}_{2}^0\pm \sqrt{3}{\cal C}_{2}^2)\sim 3y^2-r^2, \ \
3x^2-r^2,
\end{eqnarray}
Thus the degenerate ground state is formed by 
the states $|0\rangle_B$ and one of 
$(|+\rangle\mp |-\rangle)_B$.

From the consideration above, we conclude that 
a model with nearest-neighbor AFQ interaction of $\Gamma_3$ quadrupoles 
has a degeneracy with respect to different kinds of quadrupolar patterns.
In this sense, the $\Gamma_3$ antiferro-quadrupolar ordering model within the triplet state is similar to the three-state antiferromagnetic (AFM) Potts model \cite{potts}, which possesses a macroscopic degeneracy in the ground state. 
The degeneracy in our case should be broken by dipolar and/or octupolar intersite interactions.   

Experimentally, the distortion of ligands observed by X-ray or polarized neutron diffraction is $[\delta, \delta, \delta^{'}]$-type\cite{iwasa3,hao2}.
The dominance of this kind of quadrupolar moment is also indicated by the results of polarized neutron scattering\cite{hao} 
where the induced AFM moments is about 2 times larger for the field direction $(0,0,1)$ than for $(1,1,0)$. 
Taking $3z^2-r^2$-type quadrupolar field on one sublattice, however, 
the only way to keep the the local $T_h$ symmetry at the other sublattice is to break the time-reversal symmetry.  
Namely, 
if either $|+\rangle$ or $|-\rangle$ is realized,  it is possible that the lattice distortion with an $O_2^2$ component is absent.
However, a large magnetic moment should emerge in this case, in 
contradiction to the experimental results.
If  the time-reversal symmetry is preserved, the only possible  combinations are $(|+\rangle+|-\rangle)$ and $(|+\rangle-|-\rangle)$.  
In this case, a non-zero $x^2-y^2$-type quadrupolar field should emerge on one of the sublattices.  Namely, the AFQ order should always be a mixture of 
$O_2^0$ and $O_2^2$ components in the triplet subspace.
It is an interesting future problem to determine the accurate AFQ pattern in PrFe$_4$P$_{12}$ by detailed comparison of experiment and theory.

\section{Summary}

We have studied two elements which seem essential in understanding physics of Pr skutterudites: on-site hybridization and intersite interaction.
The hybridization between $4f$ electrons and the surrounding ligands plays a decisive role for the characteristic CEF splitting and the Kondo-like behavior. 
Emergence of the Kondo effect in skutterudites depends on the nature of the triplet wave functions; whether they are dominantly composed of $\Gamma_4$.
With this observation we have explained why the skutterudite PrFe$_4$P$_{12}$ shows the Kondo effect,  while another skutterudite PrOs$_4$Sb$_{12}$ does not.
More detailed account has been given in refs.\citen{Otsuki1,Otsuki2}.

As concerns typical phenomena brought about by intersite interactions, we have proposed a comprehensive picture for the $\Gamma_3$-type quadrupole order in PrFe$_4$P$_{12}$.
The low- and the high-field phases have been reproduced simultaneously
with the assumption of vanishingly small splitting between the singlet and triplet states.  
The key point for appearance of the high-field phase
is that a level crossing occurs only for the magnetic field direction $(111)$ 
in the pseudo-quartet CEF scheme.  
The magnitude of the dipole and octupolar interactions controls 
whether the high-field phase has the same (${\bf q}=(1,0,0)$) or different (${\bf q}=0$) ordering vector as that in the low-field phase.   
In the ${\bf q}=0$ case, we expect macroscopic lattice distortion in the high-field phase, and the corresponding elastic anomaly.   It is highly desired that experiment in the near future will determine the type of ordering in the high-field phase.

\section*{Acknowledgements}
We would like to thank K. Iwasa and L. Hao for showing their experimental results prior to publication, and acknowledge useful information from T. Sakakibara and T. Tayama.   This work has been supported by a Grant-In-Aid for Scientific Research from the Ministry of Education, Science, Sport and Culture, Japan.

\end{document}